\documentclass[preprint,12pt]{elsarticle}

\usepackage{graphicx,subfigure}
\usepackage{listings}

\usepackage{amssymb}

\usepackage[usenames,dvipsnames,svgnames,table]{xcolor}

\usepackage{setspace}

\journal{Journal of Theoretical Biology}
\begin{document}
\begin{frontmatter}



\title{Local behavioral rules sustain the cell allocation pattern in the combs of honey bee colonies (\textit{Apis mellifera})}


\author[label1]{Kathryn J. Montovan}
\address[label1]{Center for Applied Mathematics, Cornell University, Ithaca, New York}

\author[label2]{Nathaniel Karst}
\address[label2]{Mathematics and Science Division, Babson College, Wellesley, Massachusetts}

\author[label3]{Laura E. Jones}
\address[label3]{Department of Ecology and Evolutionary Biology, Cornell University, Ithaca, New York}

\author[label4]{Thomas D. Seeley}
\address[label4]{Department of Neurobiology and Behavior, Cornell University, Ithaca, New York}

\vfill


\end{frontmatter}


\section*{Highlights}
\begin{itemize}
\item Accepted models do not maintain the known pattern after brood cells are vacated.
\item New local behavioral rules maintain the pattern over multiple brood gestation cycles.
\item Biasing the queen's walk towards the center of the comb further stabilizes the pattern.
\end{itemize}

\pagebreak

\section*{Abstract}
In the beeswax combs of honey bees, the cells of brood, pollen, and honey have a consistent spatial pattern that is sustained throughout the life of a colony. This spatial pattern is believed to emerge from simple behavioral rules that specify how the queen moves, where foragers deposit honey/pollen and how honey/pollen is consumed from cells.  Prior work has shown that a set of such rules can explain the formation of the allocation pattern starting from an empty comb. We show that these rules cannot maintain the pattern once the brood start to vacate their cells, and we propose new, biologically realistic rules that better sustain the observed allocation pattern. We analyze the three resulting models by performing hundreds of simulation runs over many gestational periods and a wide range of parameter values. We develop new metrics for pattern assessment and employ them in analyzing pattern retention over each simulation run. Applied to our simulation results, these metrics show alteration of an accepted model for honey/pollen consumption based on local information can stabilize the cell allocation pattern over time. We also show that adding global information, by biasing the queen's movements towards the center of the comb, expands the parameter regime over which pattern retention occurs.

\section{Introduction}
\label{sec:intro}
Honey bee colonies benefit from a high degree of internal organization, but it is sometimes unclear how the thousands of bees work together to make decisions and create stable, colony-level patterns. Many studies have shown that different elements of colony-level organization and decision making rely on individual bees performing fairly simple actions. For example, we know that foraging individuals perform waggle dances that recruit others to desirable foraging locations in appropriate densities \cite{Dyer2002}, and that new colonies collectively choose the best nest cavity based on information gathered by many individual bees \cite{Seeley_etal2006}. In this paper, we consider how the actions of individual bees can cause the self-organized creation and maintenance of a colony-level storage pattern for brood, honey, and pollen in a colony's combs. 

Seeley and Morse (1976) described a general cell allocation pattern in the nests of honey bees: a dense brood clump surrounded by cells storing pollen, and with honey stored in periphery cells, mostly in the upper region of the comb \cite{Seeley:1976gb}. 
This distribution of different types of cells confers several benefits to the colony. First, it helps ensure that a colony's brood are raised at the proper temperature.  Tautz \emph{et al.} (2003) showed that the temperature at which pupae are incubated has a significant impact on their ability to perform foraging functions as adults \cite{Tautz:2003co}. Fehler \emph{et al.} (2007) connected temperature, colony efficiency, and brood density by demonstrating that brood areas with larger percentages of open cells require more attention from workers in order to maintain an optimal brood rearing temperature \cite{Fehler_etal2007}. Starks and Gilley  (1999) deepened this connection between the temperatures and brood health in their observation that that worker bees themselves act to shield brood from temperature fluctuations by positioning themselves on particularly warm areas on the interior of the hive's walls \cite{StarksGilley1999}.  Camazine (1990) argued that along with worker behavior, the physical distribution of different cell types can act to maintain proper temperature by suggesting that concentrating brood cells near the middle of the nest helps insulate the larvae from fluctuating environmental conditions \cite{Camazine:1990jh}. Thus, an advantageous positions of brood cells frees workers from needing to perform some thermoregulation tasks. 

Second, maintaining a ready supply of pollen near developing brood increases work efficiency by the nurse bees in a colony. Cralsheim \emph{et al.} (1992) showed that the primary consumers of pollen are nurse bees which feed the brood \cite{Crailsheim:1992kw}, and Camazine (2001) noted that pollen storage near brood cells would theoretically reduce the time and energy spent by nurse bees in retrieving stored pollen \cite{Camazine:2001vp}. Taken in total, the existing literature presents a convincing case for the effectiveness of a densely populated region of brood cells immediately surrounded by a ring of pollen storage cells, with honey storage cells filling the remainder of the comb. 

%

Much work has been done to understand how this pattern is created within the nest, but none of this research has considered pattern maintenance after brood begin to vacate their cells. Originally, it was believed that the pattern arises because each bee follows an internal ÒblueprintÓ, placing each product in its associated cells according to an overall plan \cite{Seeley:1976gb, Seeley:1983tm, Winston:1987we, Camazine:1991eu}. Camazine refuted this argument by observing that when empty comb is inserted into the brood region it is initially filled with both pollen and honey, but that fairly quickly these cells are emptied and filled with brood \cite{Camazine:1991eu}. This observation led to cellular automata models \cite{Camazine:1991eu, Camazine:2001vp} and simplified differential equation models  \cite{Camazine:1990jh, Jenkins:1992cc} of self-organized pattern formation in which the storage patterns result from each bee following simple behavioral rules that do not rely on global information about the nest. These models are able to explain the creation of an idealized self-organized pattern on an initially mostly-empty sheet of comb, but they only consider the first 20 days; the simulations based on these models stop before the first bees vacate their brood cells.   

A more recent model for the storage pattern developed by Johnson (2009) combines the idea of self-organization with gravity-based templates (\emph{i.e.}, blueprint-like rules) which bias the movement of nectar handlers towards the top of the comb and help produce a more realistic pattern with honey stored near the top of the comb \cite{Johnson2009}. This model includes two kinds of global information, templates for nectar storage and brood cells, but it too only considers the pattern formation before young bees start to vacate their cells (the first 20 days).

In this paper we present a cellular automata model that uses simple, local, biologically relevant rules to maintain storage patterns over multiple brood cycles. We start with the model developed by Camazine in 1991 \cite{Camazine:1991eu}, which can create a self-organized pattern on a nearly empty comb (now referred to as model 1) and change some of the rules in biologically reasonable ways to create models that both initially create and then steadily maintain the comb allocation patterns once young bees begin to vacate their cells (models 2 and 3).

Our first modification is in the implementation of a rule that specifies that consumption of nectar and pollen is brood-density dependent. This behavioral rule is based on the observation that most of the stored pollen and a good amount of the stored honey are consumed by nurse bees feeding the brood \cite{Crailsheim:1992kw}. These bees originate from the brood cells and would find nearby food cells more frequently than far away cells. In model 1 \cite{Camazine:1991eu}, when a bee is searching for a cell from which to consume nectar or pollen, the cell is chosen randomly from all of the cells in the comb, and the number of loads taken is linearly proportional to the local density of brood. Thus, when a cell is chosen, a greater number of loads are taken if there are many brood cells nearby.  We argue that this is not realistic, because nurse bees cannot carry more nectar or pollen than other bees. Instead, they are more likely to choose cells close to the brood. Thus, we propose modifying the implementation of this rule to linearly increase the probability of choosing cells near brood based on the local brood density and then take only one load each time the cell is chosen (models 2-3). Thus cells near brood are more likely to be chosen but only one load is taken from each chosen cell.

Our second modification is in the way that the queen moves (model 3). In the original model (model 1, based on \cite{Camazine:1991eu}) and model 2, each time the queen moves she choses a random direction and moves one step. We consider the option that she senses heat gradients on the surface of the comb and modifies her direction of movement based on these heat gradients. These heat gradients result from a colony-level effort to maintain an acceptable temperature for brood survival and development \cite{Fehler_etal2007, Tautz:2003co}. The workers in the colony maintain the temperature in the brood region by heating the caps of individual brood cells \cite{Bujok_etal2002}, entering empty cells within the brood region and heating adjacent brood through the cell walls \cite{Kleinhenz_etal2003}, creating evaporative cooling \cite{Heinrich1985}, and using their own bodies to make a heat-shield \cite{StarksGilley1999}. These thermoregulatory actions, focused on the brood region, can create thermal gradients across the nest \cite{HumphreyDykes2008} that are qualitatively similar to the gradients measured in colonies of bees \cite{BecherMoritz2009}. When the comb is full, as will be the case for most of our modeling, there is a well established temperature gradient from center to edge \cite{Kronenberg:1982by}. It has been shown repeatedly that bees are aware of and change their behavior in response to the temperatures that they experience \cite{GrodzickiCaputa2005, Vega_etal2011} and it is reasonable to believe that the queen can sense these thermal gradients and respond accordingly. There has been no research done on the queen's specific response to thermal gradients so we model them according to our best intuition and present this as an open question in honey bee behavior.

\section{Methods}
\label{Meth}

In a comb that has a well formed cell allocation pattern, the actions of the bees can either maintain or destroy this pattern over time. The difference between maintenance and destruction lies in the choice of parameters for key functions, as well as in implementation choices for important pieces of the model. Since a significant amount of work has already been done on the formation of the pattern, we will focus on the ability of simple rules to maintain the storage pattern over realistic timeframes of multiple brood cycles. We begin by outlining the overall structure and computational aspects of the simulator and parameter selection scheme (Section \ref{sec:implementation}). We then detail the  three main components of the models we compare: queen movement and oviposition (Section \ref{sec:queen}), nectar/pollen collection and deposition (Section \ref{sec:deposition}), and nectar/pollen consumption (Section \ref{sec:consumption}). Finally, we will confirm that the proposed rules are also able to form the pattern on a nearly empty comb. 

\subsection{Model implementation}\label{sec:implementation}

We implement the models using a cellular automata simulation model in Matlab \cite{Matlab}. The modeled comb is 45 cells wide by 75 cells tall with  hexagonal cells, which matches the approximate number of cells on one side of a full depth Langstroth frame. We simulated a season of 60 days, with a 12 hour day-night cycle. The simulation has hour-long time steps, where foragers deposit honey and pollen during the day, and bees consume honey and pollen and the queen lays eggs into suitable cells during all hours.

At the beginning of each hour, we determine the number of eggs the queen attempts to lay as she walks along the comb (see Section \ref{sec:queen}), the amount of honey and pollen deposited and consumed (see Section \ref{sec:deposition}). In order to avoid simulation artifacts caused by some tasks being preferentially performed before others, we randomize the sequencing of deposition, consumption and oviposition events each hour. Brood mature in approximately 21 days and then vacate their cells \cite{Winston:1987we}, so in the model, the 21-day-old immature bees are randomly partitioned into 24 equally sized groups (up to rounding error), one of which vacates its cells at the end of each hour. 

Unless specified, each model run was initiated with a completely full comb with an ideal pattern of a center region of brood, surrounded by a ring of pollen, and honey in all remaining cells. The assignment of type to each cell is deterministic and constant across all simulations. The brood region is a circular disk centered in the middle of the comb with radius 18 cell lengths. Around this brood region is a ring of pollen 4 cell lengths wide. The rest of the comb is filled with honey. Each storage cell has the capacity to contain up to 25 loads of honey or 15 loads of pollen. This is consistent with \cite{SchmicklCrailheim2007} and is between the estimates used in \cite{Camazine:1991eu} and \cite{Johnson2009}. The initial amount of nectar in each pollen and honey cell was chosen uniformly randomly from the ranges of $1 - 15$ loads and $1 - 25$ loads, respectively. Similarly, the initial age of each brood cell is chosen uniformly randomly from $1 - 21$ days. While developing the model, we explored multiple capacities and found that changing the capacity of the honey and pollen cells within the established ranges (pollen: $15-20$ loads per cell, honey: $20-40$ loads per cell) did not qualitatively change the resulting allocation patterns.

Other parameter estimates for this system are somewhat speculative, so we consider a wide range of values for each model parameter.  To sample the parameter space efficiently and enable analysis of model sensitivity to variation in parameter values, we used a Latin hypercube sampling structure. Latin hypercube sampling chooses $m$ equally likely values for each parameter and then randomly selects (without replacement) from these values to create a unique parameter set for each of the $m$ model runs \cite{Blower:1994ul, Ellner:2006wl}. We create 200 unique parameter sets that we use to analyze all three models. Ranges for the key parameters in the model (Table \ref{tab:param}) were chosen based on the relevant literature, with ranges extended to acknowledge uncertainty in parameter estimates. Reasoning for particular parameter choices is included in the related model sections below. 

As discussed in Section \ref{sec:intro}, we consider three models. While the details of each will be elaborated below, the main components and their similarities and differences are as follows:

\emph{Model 1:} The queen performs a random walk across the comb and attempts to oviposit in suitable cells. Workers attempt to deposit honey and pollen in cells sampled uniformly randomly from all cells. Workers attempt to consume honey and pollen sampled uniformly randomly from all cells, with the number of loads taken proportional to the number of neighboring brood cells. 

\emph{Model 2:} The queen performs a random walk across the comb and attempts to oviposit in suitable cells. Workers attempt to deposit honey and pollen in cells sampled uniformly randomly from all cells. Workers attempt to consume 1 load of honey or pollen at a time, with the probability a cell will be selected proportional to the number of neighboring brood cells.

\emph{Model 3:} The queen performs a random walk biased towards the center of the comb and attempts to oviposit in suitable cells. Workers attempt to deposit honey and pollen in cells sampled uniformly randomly from all cells. Workers attempt to consume 1 load of honey or pollen at a time, with the probability a cell will be selected proportional to the number of neighboring brood cells.

When mathematically defining the exact mechanisms by which these extractions, depositions and ovipositions occur, it will be convenient to have symbols to refer to certain classes of cells. At every time $t$, we can partition the cells of the comb into four subsets: $E(t)$, the empty cells; $H(t)$, cells containing honey; $P(t)$, cells containing pollen; $B(t)$, cells containing brood. We define $N = 45 \times 75 = 3375$ to be the total number of cells on the comb.

\subsection{Queen movement and egg laying}\label{sec:queen}
In order to capture the variability in the queen's walk across the comb, we use one of two probability distributions to model the direction of her movement: 
\begin{itemize}
\item[(1)] uniform distribution on the interval $[-\pi,\pi]$ (random walk)
\item[(2)] wrapped Gaussian distribution with mean $\theta$ and standard deviation $\sigma$ on $[-\pi,\pi]$ (biased random walk). 
\end{itemize}
In both cases, the mean $\theta = 0$ represents the angle pointing the queen from her current position towards the center of the comb. Once a direction is chosen, the queen moves to the nearest cell in that direction. 

While the uniform angle distribution is relatively straightforward, the exact mechanism by which we introduce a bias in the queen's movements towards the center of the comb is important. For simplicity, we implement an affine scaling of the standard deviation of the distribution as a function of the queen's distance from the center of the comb of the form 
\begin{eqnarray}
\sigma = \sigma_o - (\sigma_o - \sigma_c) {d \over d_{max}},
\end{eqnarray}
where $d$ is the distance in cell lengths from the center of the cell on which the queen is currently located to the center of the cell in the center of the comb, and $d_{max}=\sqrt{22^2+37^2}$ is the maximum distance from any cell to the center of the comb. The tunable parameters $\sigma_o$ and $\sigma_c$ describe the desired standard deviation when the queen is located at the center (origin) and corners of the comb, respectively. With $\sigma_o$ sufficiently large, the wrapped Gaussian produces nearly uniformly random angles when the queen is near the center of the comb. If $\sigma_o > \sigma_c$, the queen's movement becomes increasingly biased towards the center of the comb as she moves farther away from it. We set $\sigma_0 = 5$ and $\sigma_1 = 2.828$. With this choice, the queen visits cells at the edge of the comb roughly half as many times as cells near the center of the comb. 


The number of cells visited by the queen in one hour, $n$, is determined by the Latin hypercube sampling for each model run and is between 60 and 120 cells per hour. This parameter range was chosen because the queen lays between 1000 and 2000 eggs in a day, with is equivalent to $42-84$ eggs per hour \cite{Nolan:1925uy, Bodenheimer:1937vk, Camazine:1991eu}. We selected a range from $60-120$ cells visited per hour because many attempts to lay eggs fail either because the cell is already in use or because it is too far from the nearest brood cell. In an empty comb the queen will lay roughly the desired maximum number of eggs and in a more full comb her efficiency decreases as she spends more time searching for suitable cells. The set of cells $A(t)$ which the queen finds acceptable are empty and within radial distance $r_b$ from a brood cell. In symbols, 
\begin{eqnarray}
A(t) = \{e~:~ e \in E(t), \min_{b \in B(t)} d(e,b) \leq r_b\},
\end{eqnarray}
where $E(t)$ is the set of empty cells at time $t$, and $d(x,y)$ is the Euclidean distance measured in cell lengths between the center of cell $x$ and the center of cell $y$. This distance threshold $r_b$ is varied in the Latin hypercube sampling design between 1 and 4. The upper end of this range was chosen to match Camazine's cellular automata model \cite{Camazine:1991eu} and the shorter distances test the sensitivity of the models to this parameter. 

 \subsection{Nectar and pollen collection and deposition}\label{sec:deposition}
 \label{sec:deposition}
Both honey and pollen are deposited into cells which are empty or partially filled with the same substance as is being deposited. Pollen foragers and honey storers examine multiple cells when depositing loads of food \cite{Calderone:2002go} and deposit less honey and pollen when the comb is full \cite{Seeley:1989gj}. To be consistent with these observations, each forager selects cells uniformly randomly from \emph{all} comb cells and is allowed up to 6 attempts to find a suitable cell. This modeling choice serves two purposes. First, workers do not need to have global information about the location of all honey and/or pollen cells at a given time as they would if the cells were chosen randomly from the available honey/pollen cells and empty cells. Second, a worker aborting the search for an appropriate cell on this comb approximates the worker going to find an empty cell on another comb when the simulated comb is becoming overly full. This creates the random deposition with the desired decreased deposition rate for full combs. We note that this interpretation conforms to the descriptions in Camazine \cite{Camazine:2001vp}.

In order to describe deposition in our agent-based model, we must describe the collection and deposition rates in terms of actions of individual bees. We calculate the number of individual loads of honey and pollen that are deposited into the comb each hour from established yearly totals and measured bee load capacity. A typical colony collects $60$ kg of honey in a season, with $40$ mg of honey in each load, $180$ days in the summer season, and approximately 10 sheets of comb per colony \cite{Camazine:1991eu, Camazine:2001vp}. This results in approximately $833$ loads on average entering each sheet of comb in the hive every day. This estimate was then increased to account for the fact that, in our models, many attempts to deposit honey are unsuccessful. For each model run, the average number of loads of honey collected per hour $\omega$ was determined by uniform sampling between 1000 and 4000 loads per day for the Latin hypercube setup. The average number of loads of pollen collected per hour is chosen by Latin hypercube sampling as a fraction of the collected honey, $\rho_{ph}\in [.2,1]$ so the total amount of pollen collected in a season is $\rho_{ph}\omega$. The ratio $\rho_{ph}$ has been observed to be about $0.26$ \cite{Camazine:1991eu}. Our model extends this range to look at the sensitivity and consider potential changes in storage ratios for the full comb within the simulated nest. 

Pollen and honey availability depends on seasonally variable flowers and weather dependent favorable foraging conditions. To capture this, we consider three different types of temporal variability in nectar foraging, with the method chosen at random for each model run. The amount of honey and pollen collected per day are either 
\begin{itemize}
\item[(1)] constant in time and equal $\omega$ and $\omega \rho_{ph}$, respectively.
\item[(2)] drawn uniformly randomly from $[0,2\omega]$ and $[0,2\omega \rho_{ph}]$, respectively.
\item[(3)] subject to a Markov process in which the amount of honey and pollen collected are either identically zero or equal to $2\omega$ and $2 \omega \rho_{ph}$, respectively, with probability $0.70$ that the amount collected on a given day will be the same as the amount collected the day before. 
\end{itemize}

The transition probability in the Markov process model was chosen to create realistic fluctuations in food availability. This could be refined, but for our model we decided to keep this element fairly simple. In all of these cases, the total amount of food collected during the modeled season is set to the predefined amounts for each type of food. The daily amounts were then used to calculate the hourly collection rates which is simply one twelfth of the daily collection rates. 

\subsection{Nectar and pollen consumption}\label{sec:consumption} Food consumption is modeled by randomly choosing a cell in the comb and taking a load out of this cell if it contains the desired food type. We assume that consumption depends heavily on the number of nearby brood. The dimensionless brood density within distance $r_n$ at a cell $c$ is given by
\begin{eqnarray}
D_{r_n}(c) = {|\{b~:~b \in B(t) , d(b,c) \leq r_n\}| \over  3 r_n(r_n+1)},
\end{eqnarray}
where $B(t)$ is the collection of brood cells at time $t$, and $d(b,c)$ is the Euclidean distance measured in cell lengths from the center of cell $b$ to center of cell $c$. The denominator is given by the observation that the total number of cells whose centers are within $r_n$ cell lengths a the center of a given cell on a hexagonal grid is $6 + 12 + \ldots + 6r_n = 6(1+2+\ldots+r_n) = 6r_n(r_n+1)/2$ (excluding the chosen cell itself). 

The brood density dictates honey and pollen consumption in all models considered. In model 1, cell choice is uniformly random and the number of loads of nectar taken from a selected cell $c$ is linearly dependent on the local brood density $D_{r_n}(c) $. 
\begin{eqnarray}
P(\mbox{cell $c$ chosen})  & = &  {1 \over N} \\
n_L  & = &  \min(\mbox{loads left}, \lfloor 1 + D_{r_n}(c)  (k-1) \rceil).
\end{eqnarray}

In models 2 and 3, cell choice is linearly proportional to the local brood density, and the number of loads of honey or pollen taken from a selected cell is constant. 
\begin{eqnarray}
P(\mbox{cell $c$ chosen})  & \propto &  1 + D_{r_n}(c)  (\ell-1) \\
n_L  & = &  1.
\end{eqnarray}
In all models, cell choice for honey or pollen removal is taken over all comb cells, regardless of whether a cell is (partially) filled with the desired type of food or not. If the desired type is not found in the chosen cell, then another cell is chosen, with up to six cells being checked before the process is abandoned and the model moves on to the next task. Note that in both methods workers do not need to have global information as to the location of all honey and/or pollen cells at a given time.

Camazine originally set $r_n = 1$ and $k = 10$ \cite{Camazine:1991eu}. Here we have expanded these definitions to $r_n \in [1,4]$ and $k  \in [5,20]$ in order to determine the sensitivity of each model to these parameters. 

The amount of honey and pollen consumed over the entire modeled season is calculated as a ratio of amount foraged. Consumption is assumed to be constant throughout the season, and during all hours of the day. The ratios of pollen and honey consumption to collection ($\rho_p$  and $\rho_h$, respectively) were chosen to be in the range of $0.9-1.1$ since our interest is in pattern maintenance after the comb fills. This range allows us to consider the phase when much of the incoming honey is being deposited in other non-brood combs. Within the nest, central combs contain brood and other combs are mostly used for the storage of honey  \cite{Seeley:1976gb}. Colonies have mechanisms that ensure that foraging does not exceed available storage capacity, which include comb building and colony splitting to create a new colony \cite{Winston:1987we}. These mechanisms, combined with the use of a small number of combs for brood, should maintain the rate of incoming honey and pollen to these brood combs to, on average, replace the consumed honey and pollen. Thus on these combs, we expect to see ratios of consumption to collection close to $1$ after the comb is full and the pattern is established. Otherwise, in time, the comb would become either overfull or completely empty.

\subsection{Brood and Pollen Ring metrics} 

To assess the level of pattern retention during the simulation runs, we developed two metrics that describe the compactness of the brood region and the level of definition of the pollen ring (or gap of empty cells). The brood metric is the average number of adjacent brood for each brood cell. 
\begin{eqnarray}
m_b(t) = {1 \over |B(t)|} \sum_{b \in B(t)} |\{x ~:~ x \in B(t), 0 < d(x,b) \leq 1\}|, \label{eqn:bc}
\end{eqnarray}
where $B(t)$ is the collection of brood cells at time $t$, and $d(x,b)$ is the Euclidean distance measured in cell lengths from cell $x$ to cell $b$. Note that $m_b(t)$ is undefined if $|B(t)| = 0$, that is, if there are no brood on the comb at time $t$. We observed qualitatively that in simulations with brood compactness metric $m_b \geq 5.25$, the brood cells are sufficiently dense to fit the observed pattern. 

The pollen metric is the average distance from each honey cell to the nearest brood cell, \emph{i.e.}, the smallest number of cells visited when traveling from a honey cell to the nearest brood cell. 
\begin{eqnarray}
m_p(t) = {1 \over |H(t)|} \sum_{h \in H(t)} \min \{d(b,h)~:~b\in B(t)\}, \label{eqn:pr}
\end{eqnarray}
where $H(t)$ is the collection of cells containing honey at time $t$, and $d(b,h)$ is the Euclidean distance measured in cell lengths from the center of cell $b$ to the center of cell $h$. Note that $m_p(t)$ is undefined if $|H(t)| = 0$, that is, if there are no cells storing honey on the comb at time $t$.  In this case we observed that pollen metric $m_p \geq 12$ indicates a well-formed pollen ring, \emph{i.e.}, one that forms a strong separation of honey cells from brood cells. In combination, these two metrics accurately describe how well the allocation adheres to the desired pattern. We use these metrics to assess the sensitivity in the model predictions over a range of reasonable parameter values. 

\subsection{Pattern Formation} In addition to testing pattern maintenance, we investigate the ability of each model to create the desired pattern on a nearly empty comb (similar to \cite{Camazine:2001vp}). We perform the same simulations as above, but now set the initial comb storage pattern to be mostly empty with a clump of 7 brood cells (one brood cell with 6 adjacent brood cells) in the center of the comb. The parameter value ranges for some parameters were adjusted for the formation phase. We considered 100 parameter sets with the radius for the brood requirement ($r_b$) restricted to 2-4 since radii of 1 resulted in no new brood in the full (pattern sustenance model). We also restricted the ranges on the ratio of pollen collected to honey collected ($\rho_{ph}\in(.21,.45)$), the expected ratio of pollen consumption to pollen collection ($\rho_p\in(.9,1.08)$), the expected ratio of honey consumption to honey collection ($\rho_h\in (0.49,0.69)$, and the preferential consumption pressure near brood cells ($k \in (5,15)$). These adjusted and narrower parameter ranges helped us look at the pattern formation locally near measured parameter estimates when the comb is being filled early in the season or in a new nest.

\subsection{Sensitivity testing}\label{sec:sensDesc}
We performed a global sensitivity analysis to assess the relative impact of each parameter on pattern retention for each of the 3 models. Because parameter estimates are uncertain, parameter ranges included maximum values up to a factor of 5 times larger than baseline values. For a complete list of parameters and their ranges, see Table \ref{tab:param}. As Latin hypercube sampling is an efficient way of sampling a large parameter space and our model is computationally intensive,  we employed the same multidimensional hypercube used for parameter determination in our sensitivity testing  \cite{McKay:1992LHS, Blower:1994ul}.   For each model scenario, 200 randomized parameter sets were generated by our hypercube. For each of these, we simulated 60 days;  model metrics were then computed for days $20-60$, and the values averaged. This time window includes multiple brood gestation periods,  but omits transient dynamics due to comb initialization. We discard any run in which the brood clumping or pollen ring metric were undefined at any time between day 20 and day 60, leaving $N_1$, $N_2$ and $N_3$ viable runs for models 1, 2, and 3 respectively. Recall that the brood clumping and pollen ring metrics are undefined when there are no honey cells and no pollen storage cells, respectively, on the comb. These scenarios can occur, for instance, if the ratio of honey collection to consumption $\rho_h$ is relatively large and relatively small, respectively. 


After preprocessing the data, we scale each parameter value so that it is a percent of the observed parameter range in the $N_i$ simulation runs, with 0.00 representing the minimum value and 1.00 representing the maximum value. We then perform multiple linear regression on the scaled data. We discard the intercept information from the linear regressions for both the pollen ring and brood region metrics, but note that the inclusion of this information in the regression is critical; without it, the least-squares method will produce a linear function for which each metric is equal to 0 when all parameters are equal to zero which is clearly not appropriate in the system modeled here.

We must interpret the remaining components, the so called elasticities of the metrics, with the scalings we have performed in mind. An elasticity value of 2 indicates that increasing the corresponding parameter from the bottom of its range to the top of its range increases the metric by 2 on average. (Notice that the the average metrics have not been scaled, so they should not be interpreted as percentages.) Similarly, an elasticity value of -3 indicates that increasing the corresponding parameter from the bottom of its range to the top of its range decreases the metric by 3 on average. 


\section{Results}
\label{Res}
Our results show that the cell allocation patterns of brood, pollen, and honey can be maintained over multiple brood gestation cycles by simple behavioral rules. To compare pattern retention across the 3 models, we first generated 200 parameter combinations using Latin hypercube sampling over the parameter ranges featured in Table \ref{tab:param}. For each parameter combination, 3 separate 60 day simulations of the comb were completed, one for each model, with the initial state of each simulation being the ideal cell allocation pattern described in Section \ref{sec:implementation}. 

\emph{Comb snapshots:} We begin by examining the cell allocation pattern across the comb. In Figures 1 -- 3, we plot the comb at several points in time for models 1 -- 3, respectively. The simulation run featured in each figure maximized the product $\overline{m_p} \cdot\overline{m_b}$, where $\overline{m_p}$ and $\overline{m_b}$ are the pollen ring and brood clumping metrics, respectively, averaged over days 20 -- 60. While this product of averaged metrics is just one way we might define good performance in a simulation, we have found that it is a good indicator of pattern retention. Moreover, it is simple both in its form and its computation. We average only after the first 20 days so as to not allow the initial ideal pattern to unduly influence the value of the averaged metrics. 

Monitoring the simulations at days 20, 40, and 60 in Figures 1 -- 3 allows us to easily compare across all 3 models. In Figure 1, we see that model 1 has a fairly well defined and compact brood region at day 20. By day 40, this region has deteriorated, with the brood being both more diffuse and with honey storage cells occurring more frequently in the brood region. At day 60, the region containing brood has expanded considerably and is fair less densely populated with brood cells than at the previous snapshots. Pollen and honey storage cells are intermixed throughout the comb. 

We compare these trends to the behavior of model 2 in Figure 2. Here we observe a compact brood region and well defined pollen ring across the 60 days of simulation. Some honey storage cells do encroach on the brood region, but most of these are converted to brood cells between snapshots, indicating the phenomenon is transient. A likely scenario is that the cell containing honey was recently vacated by an immature bee; due to preferential removal, this cell stays empty or almost empty much of the time, which increases the probability that the queen will lay an egg in it when she is next at the cell. Figure 3 shows that model 3 produces qualitatively similar results to model 2. 

\emph{Metric time series comparison:} In order to tease apart quantitative differences in qualitatively similar patterns, we plot the pollen ring and brood clumping metrics over time in Figure 4. Here we plot only the 20 simulations runs for each model that maximize the product $\overline{m_p} \cdot\overline{m_b}$, where again $\overline{m_p}$ and $\overline{m_b}$ are the pollen ring and brood clumping metrics, respectively, averaged over days 20 -- 60. Under model 1, the brood clumping metric seems to stabilize up to probabilistic fluctuations after day 20, albeit it to a value that is below our threshold of $m_b(t) = 5.25$. This agrees well with our observations in Figure 1: model 1 produces a brood region up to and including day 60, but the region is relatively diffuse. Most traces of the pollen ring metric are monotonically decreasing up to probabilistic effects even up to day 60. This is good agreement with our results in Figure 1, as we noted that the diffuse brood region is increasingly infiltrated by honey storage cells. We note that one trace of the pollen ring metric exhibits a wild swing from low to high over the course of 20 days. In this simulation, the comb contains no honey storage cells from day 11 to day 14. When a few honey storage cells begin to appear day 14, they are at first quite close to the brood patch, but as new groups of brood begin to emerge, the average distance between the few honey cells and the remaining immature brood cells begins to grow quite quickly, leading to a spuriously high metric. As the vacated brood cells begin to fill with honey at approximately day 40, the pollen ring metric begins to decrease to a more reasonable range, both because these new honey storage cells are relatively close to the brood cluster, and because a larger number of honey storage cells implies that outliers contribute less to the average minimum distance from honey to brood. This example and others like it motivate us to disregard any simulation run which at any point has an undefined pollen ring or brood clumping metric, as the metrics of these simulations cannot be trusted to convey accurate information about the retention of the pattern.

\emph{Metric space comparison:}  Note that Figure 4 shows both models 2 and 3 exhibit brood clumping and pollen metrics that are relatively constant and above their respective thresholds from day 20 to day 60. As there is relatively little change over time in the brood clumping and pollen metrics, we can sacrifice the temporal component of the data in Figure 4 and plot each of the 20 runs for each model in metric space as seen in Figure 5. Here we plot the average brood clumping metric versus the average pollen ring metric for all simulation runs which do not have undefined pollen ring or brood clumping metric at any time between day 20 and day 60 for each of the 3 models. All averages are performed over the interval from day 20 to day 60. Similar to Figure 4, a point in a gray region in Figure 5 represents a simulation run in which one or both of the metrics averaged from day 20 to day 60 was below threshold. We note that although the results in Figure 4 might lead us to expect  that there are simulation runs of model 1 in which the average brood clumping metric between day 20 and day 60 is above threshold (and similarly for the pollen ring metric), Figure 5 indicates that no simulation run of model is above threshold with respect to both averaged metrics. After introducing the revised honey/pollen rule in model 2, we observe 9 simulation runs that are above threshold with respect to both metrics. The parameter combinations leading to this outcome are listed in Table \ref{tab:model2}. If in addition we bias the queen's movement towards the center of the comb as in model 3, we observe 16 simulation runs that are above threshold with respect to both metrics. The parameter combinations leading to this outcome are listed in Table \ref{tab:model3}. The significance of the parameter combinations that lead to pattern retention in models 2 and 3 will be discussed in Section \ref{sec:discussion}.

Figure 5 seems to indicate that there is a significant difference between the mean of the time-averaged pollen ring and brood clumping metrics of model 1 and the corresponding means of models 2 and 3. We can quantitatively confirm this intuition by performing one-way ANOVA. The results for the test applied to the time-averaged pollen ring metric are seen in Figure \ref{tab:anova}. We preprocess the data by removing every run in which the pollen ring metric was undefined at any time between day 20 and day 40. Recall that the pollen ring metric is undefined when there are no honey storage cells on the comb. We conclude form the results of the ANOVA test that it is highly unlikely that the observed pollen ring metrics from models 1, 2, and 3 are drawn from distributions with the same mean. A multicomparison test shows that the mean time-averaged pollen metric of model 1 is significantly different than that of models 2 and 3, while the mean time-averaged pollen metric of models 2 and 3 are not significantly different. We performed an identical analysis for the brood clumping metric and found an even more pronounced difference between the models.


\emph{Sensitivity testing:} As discussed in Section \ref{sec:sensDesc}, the use of Latin hypercube sampling for parameter selection enables us to perform a straightforward sensitivity analysis via multiple linear regression.  A graphical summary of this analysis is seen in Figure 6. Here we include only the analysis of models 1 and 2, because the sensitivity profiles of models 2 and 3 are qualitatively similar. 

For both models 1 and 2, the brood clumping metric is relatively inelastic with respect to most parameters, with the notable exception in both models being the brood clumping metric's dependence on $n$, the number of oviposition attempts per hour. In both models, increasing the number of oviposition attempts per hour increases the average brood clumping metric. This agrees with our intuition, as increasing the number of oviposition attempts per hour increases the likelihood that the queen will be place an egg in a recently vacated brood cell. 

The elasticity of the pollen ring metric varies quite widely between model 1 and model 2. For most parameters, an identical increase in parameter value in model 1 and model 2 will on average simply result in a larger decrease in the pollen ring metric in model 1 than in model 2. However, there are cases where identical parameter increases will result in an increase in pollen ring metric in model 2 and a decrease in pollen ring metric in model 1. Perhaps most notable is the parameter $k$. Recall that in model 1, the parameter $k$ represents the number of loads of honey/pollen that will be removed from a cell completely surrounded by brood if it is chosen for consumption, while in model 2, the parameter $k$ represents the ratio of the probability that a cell completely surrounded by brood will be chosen for consumption to the probability that a cell with no brood neighbors will be chosen for consumption. The elasticities of $k$ in models 1 and 2 are markedly different. Increasing $k$ in model 1 leads to a substantial decrease in the pollen ring metric, while increasing $k$ in model 2 results in a moderate increase in the pollen ring metric.

Figure 6 also features some results that might seem at first counterintuitive. For instance, increasing $r_b$, the upper bound on the minimum distance from a brood cell at which the queen will oviposit, results on average in a decrease in both brood clumping metric in both model 1 and model 2. While it might at first seem that larger $r_b$ would result in a denser brood region, our results here show that larger $r_b$ more often allows the queen to oviposit well outside the current brood region, thus lowering the brood clumping metric. With that being said, many simulation runs with $r_b = 1$ eventually had no brood on the comb, and so were not included in this sensitivity analysis. This is especially relevant in model 1, where all simulations with $r_b = 1$ eventually had no brood on the comb. Together, these illuminate a natural tension: there is a parameter threshold below which the patten disintegrates, but on average increasing the parameter decreases one or both of the parameter metrics. It bears remembering that the elasticities featured in Figure 6 are simply linear fits over the entire observed parameter range. We allow, and expect, the parameters to have nonlinear effects on the metrics that are not captured by this sensitivity analysis, as well.



\emph{Model validation:} While our work here is primarily focused on pattern retention over multiple brood gestation cycles, it is also important to confirm that the models investigated here are capable of forming the cell allocation pattern from a nearly empty comb. Our modeling framework contains the same general pattern formation processes that Camazine described \cite{Camazine:1991eu}, but to check that our models would in fact create the initial pattern of a compact brood  region surrounded by a ring of pollen, we performed simulations starting with an empty comb for all three models for the first twenty days. Figure 7 shows an example of a well formed pattern for each model at day 20. For model 1, our simulations reproduced Camazine's results \cite{Camazine:1991eu}. All three models are able to form the initial pattern for a range of parameter values. The final pattern is not perfect, but the compact brood region forms and the pollen ring is visible.

Given the stochastic nature of the simulation, there is the natural question as to whether a given simulation of a particular parameter combination is representative of the behavior in general. Figure 8 shows metric traces for 20 simulations of one parameter combination applied to models 2 and 3. The traces of both metrics are relatively tight, and in particular, all traces are qualitatively similar in that all exceed the metric thresholds most of the time. All brood metrics were within 15\% of the mean, and all pollen ring metrics were within 5\% of the mean.


\section{Discussion}
\label{sec:discussion}


This study is the first to consider how cell allocation patterns are maintained over multiple brood gestation cycles. We acknowledge that our work here is a drastic simplification of what is a rich and complex natural phenomenon, and that the patterns created and maintained by the models we present in many cases capture only some of the qualitative aspects of cell allocations observed in the wild. For instance, we ignore the existence of other combs in the colony, the highly complex and variable availability of nectar and pollen, the extreme shifts in colony population over the course of a season, anisotropies introduced by gravity, and myriad other effects. Yet it is exactly this extreme simplification that makes our results here interesting; pattern formation and retention, at least in a qualitative sense, are achievable with only a few simple rules. Below we present the level to which pattern retention occurs in each of the there models and discuss the significance of the rules and information that were necessary to introduce in order to achieve a given level of pattern retention. 

Figures 1 -- 3 show anecdotally that model 1 is not capable of maintaining the pattern over a 60 day period, while models 2 and 3 are. In order to more precisely discuss the quality of an observed pattern, we have introduced a brood clumping metric $m_b(t)$ in Equation \ref{eqn:bc} and a pollen ring metric $m_p(t)$ in Equation \ref{eqn:pr}. We average these metrics over days 20 to 60 to form $\overline{m_p}$ and $\overline{m_b}$, respectively, in order to discard transients and smooth out stochastic effects. Through observations of well formed patterns and the associated brood clumping and pollen ring metrics, we say that a simulation with $\overline{m_b} \geq 5.25$ and $\overline{m_p} \geq 12$ exhibits a well formed pattern. 

Our thresholding scheme agrees with our anecdotal evidence. Of the 200 simulation runs of model 1 performed, none exhibited a well formed pattern. To make model 2, we modify model 1 to by changing the honey/pollen consumption rules as described in Section \ref{sec:consumption}. Here we observe 9 of the 200 simulation runs exhibit a well maintained pattern. We emphasize that rules for oviposition, honey/pollen consumption, and honey/pollen deposition in model 2 are based solely on local information available to each bee. The 9 parameter combinations that resulted in a well formed pattern in model 2 are listed in Table \ref{tab:model2}. It is tempting to interpret the data listed there as definitive indicators of the types of parameter combinations that are amenable to model 2 maintaining the desired pattern. But we take any such interpretation with a grain of salt, as our sampling is probabilistic in nature and our parameter space is very large. With this warning in mind, we can make several observations. We note that $n$, the number of oviposit attempts made by the queen in an hour, is between 70 and 119. The observed minimum here is roughly 20\% higher than the minimum allowable value, perhaps indicating that small values of $n$ lead to poor pattern retention. This would agree well with the sensitivity analysis of model 1 featured in Figure 6(b). Similarly, the parameter $r_b$, the maximum distance from an existing brood cell at which the queen will oviposit, never assumes value $r_b = 1$. Here we can be more definitive, because each model 1 simulation in which $r_b = 1$ results in a comb without brood cells at some time between day 20 and day 60. The parameter $\omega$, representing the number of loads of honey/pollen collected per day, achieves a large portion of its range, as do parameters $\rho_{ph}$, $\rho_p,$ and $\rho_h$. Interestingly, the nectar collection schedule indictor $\chi$, never assumes value $\chi = 2$ which would indicate that honey/pollen collection was subject to a Markov process. This may indicate that model 2 is not capable of maintaining the pattern in the presence of such variability. Finally, parameter $k$, representing the strength of preferential choice of honey/pollen cells near brood, does not assume values in the bottom 25\% of its allowable range, perhaps indicating that pattern retention fares better when stronger preference is given to cells near brood. This is in good agreement with the sensitivity analysis featured in Figure 6(b). 

In model 3, we incorporate the preferential consumption rule of model 2 and additionally bias the queen's random walk towards the center of the comb as described in Section \ref{sec:queen}. While we have presented literature that details several behaviors of honey bees perform in response to temperature and temperature gradients, there has been, to our knowledge, no work on the effect of temperature gradients on the queen's walk. Thus, while we may speculate that the queen may be using such gradients to inform her movements across the comb and hope that empiricists investigate this hypothesis, we must for the time being treat the queen's biased random walk as an introduction of \emph{global} information into the model and acknowledge that this introduction makes pattern formation and retention somewhat less impressive. 

We observe that 16 of the 200 simulation runs of model 3 exhibit a well formed pattern. These parameter combinations are listed in Table \ref{tab:model3}. As with model 2, the parameter $n$, the number of oviposit attempts made by the queen in an hour, does not assume any value in roughly the bottom 20\% of its allowable range. Note that in contrast to model 2, the parameter $r_b$, the maximum distance from an existing brood cell at which the queen will oviposit, assumes values in its allowable range. Similarly, the nectar collection schedule indicators $\chi$ and  $k$, representing the strength of preferential choice of honey/pollen cells near brood, now assume values in their full range. In all, these expansions in parameter ranges which result in a well formed pattern together seem to indicate that pattern retention is more robust in model 3 than in model 2.

Much of our current understanding of self-organization in biological systems is on the emergence of global order from initial disorganization through local interactions between individuals. Our work extends this conversation to consider the additional requirements for maintaining order after it has been established. In some systems, maintenance could reasonably be expected from any process which can create order, but in honey bees, the rules change fairly significantly after the initial pattern formation phase and make it more difficult to maintain the pattern than to form it on an empty comb. We hope this work opens a larger discussion about whether the local interactions maintaining order are the same as those that initially allowed for self-organization, or whether new mechanisms must be investigated. 

\section*{Author Contributions:}
K.J.M. and N.K. contributed equally to develop the model, analyze the results, and write the manuscript. L.E.J. contributed to the Latin Hypercube design and sensitivity analysis. T.D.S. provided expertise on the biological details that shaped the models.

\section*{Disclosure Statement:}
None of the authors had any actual or potential conflicts of interest. 

\section*{Acknowledgments}

Kathryn Montovan and Nathaniel Karst were partially supported by
National Science Foundation Integrated Graduate Education and Research Traineeships. 



\singlespace

\section{References}



\bibliography{Comb_Allocation.bib}






\pagebreak
\clearpage

\begin{table}[!h]
\begin{tabular}{lp{7cm} cl}
\hline
Parameter  &  Description  &  Estimate  &  Range \\
\hline 
$n$  &  Queen's cell visitation rate (cells per hour)  & 60 \cite{Camazine:1991eu} &  60 -- 120 \\ 
$r_b$  &  Brood requirement radius (cells)  &  4 \cite{Camazine:1991eu} & 1 -- 4 \\
$r_n$  &  Preferential nectar consumption radius (cells)  &  4 \cite{Camazine:1991eu} & 1 -- 4 \\
$\omega$  &  Average honey collection (loads per day)  &  833 (Sec. \ref{sec:deposition})  & 1000 -- 4000 \\
$\rho_{ph}$  &  Ratio of pollen collection to honey collection   (dimensionless)  &  0.21 \cite{Camazine:1991eu}  & 0.2 -- 1.0 \\
$\rho_{p}$  &  Ratio of pollen consumption to pollen collection  (dimensionless)   & 0.99 \cite{Camazine:1991eu}  & 0.9 -- 1.1 \\
$\rho_h$  &  Ratio of honey consumption to honey collection   (dimensionless)  &  0.59 \cite{Camazine:1991eu} & 0.9 -- 1.1 \\
$\chi$  &  Temporal distribution of daily nectar and pollen collection: uniform constant ($\chi = 0$), uniform random ($\chi = 1$) and Markov clumped random ($\chi = 2$)  &  NA  &  0 -- 2\\
$k$  &  Model 1: Ratio of honey/pollen taken from cells fully surrounded by brood cells to honey/pollen taken from cells with no brood neighbors   (dimensionless)  &  10 \cite{Camazine:1991eu}  &  5 -- 20 \\
$k$  &  Models 2 and 3: Ratio of probability that a cell fully surrounded by brood cells is chosen for nectar consumption to the probability that a cell with no brood neighbors is chosen   (dimensionless)  &  10 &  5 -- 20 \\
\hline
\end{tabular}
\caption{Parameters used in simulations of models 1- 3 and the sensitivity analysis. The estimates from the literature were used as a starting point for parameter ranges. The reasoning for the given ranges based on the literature estimates are given within the relevant model description sections. For example, for queen cell visitations, the estimate is for the number of eggs laid per hour, so we inflated it to account for the queen rejecting cells, then extended the range for sensitivity testing. Similar reasoning explains the elevated range for $w$. The values for $\rho_{ph}, \rho_p, \rho_h$ apply most directly to the pattern formation phase of colony development, and were modified for the full comb.}\label{tab:param}
\end{table}

\begin{table}
\begin{tabular}{|l|l|l|l|l|l|l|l|l|l|l|l|}
\hline
Run No. & $n$ & $r_b$ & $r_n$ & $\omega$ & $\rho_{ph}$ & $\rho_p$ & $\rho_h$ & $\chi$ & $k$ & $\overline{m_b}$ & $\overline{m_p}$ \\ \hline
1&108&3&1&1420&0.5&1.07&1.05&1&9&5.32&12.5\\\hline
2&97&2&3&1945&0.76&0.99&1.08&1&19&5.29&12.51\\\hline
3&70&4&3&3370&0.33&0.94&1.09&0&18&5.3&12.84\\\hline
4&75&2&3&2275&0.67&1.08&0.92&0&14&5.26&12.26\\\hline
5&116&3&1&3520&0.81&1.06&1.04&0&16&5.38&12.22\\\hline
6&112&2&1&3445&0.73&0.99&0.95&0&13&5.29&12.29\\\hline
7&88&3&1&1210&0.32&1.06&0.99&1&9&5.3&12.72\\\hline
8&119&4&4&1165&0.57&1.03&1&1&19&5.33&12.03\\\hline
9&97&4&1&1150&0.28&0.94&1.1&0&15&5.34&12.94\\\hline
\end{tabular}
\caption{Model 2 parameter contributions that result time averaged metrics $\overline{m_b}$ and $\overline{m_p}$ that are both above their respective thresholds.}\label{tab:model2}
\end{table} 

\begin{table}
\begin{tabular}{|l|l|l|l|l|l|l|l|l|l|l|l|}
\hline
Run No. & $n$ & $r_b$ & $r_n$ & $\omega$ & $\rho_{ph}$ & $\rho_p$ & $\rho_h$ & $\chi$ & $k$ & $\overline{m_b}$ & $\overline{m_p}$ \\ \hline
1&102&4&4&1885&0.94&0.96&1.03&0&12&5.4&12.22\\\hline
2&90&4&2&1930&0.51&0.96&1.08&0&16&5.29&12.88\\\hline
3&90&3&1&1000&0.97&1.02&0.99&0&16&5.41&13.07\\\hline
4&110&3&3&2140&0.88&0.92&0.97&0&15&5.43&12.85\\\hline
5&108&3&2&3580&0.22&1.05&1.02&0&14&5.26&12.1\\\hline
6&118&4&2&2440&0.36&0.9&1.01&2&14&5.43&12.66\\\hline
7&89&2&3&1300&0.25&1.03&1.1&0&9&5.49&12.5\\\hline
8&103&3&1&1015&0.68&1.06&0.92&2&9&5.38&12.27\\\hline
9&112&2&1&3445&0.73&0.99&0.95&0&13&5.47&12.06\\\hline
10&88&3&1&1210&0.32&1.06&0.99&1&9&5.32&13.08\\\hline
11&79&4&3&3790&0.24&0.99&0.91&1&19&5.25&12.01\\\hline
12&104&2&2&3415&0.41&0.98&0.93&1&17&5.31&12.24\\\hline
13&70&2&2&2830&0.64&0.91&1.01&2&6&5.49&12.2\\\hline
14&105&3&1&3865&0.37&0.95&1.09&1&9&5.38&12.88\\\hline
15&101&1&4&3220&0.97&1&1.05&0&17&5.44&12.25\\\hline
16&99&1&3&1585&0.95&1.02&1.02&2&9&5.32&12.39\\\hline
\end{tabular}
\caption{Model 3 parameter contributions that result time averaged metrics $\overline{m_b}$ and $\overline{m_p}$ that are both above their respective thresholds.}\label{tab:model3}
\end{table}

\begin{table}
\begin{tabular}{llllll}
\hline
Source & DF & Sum of squares & Mean squares & F ratio & F probability \\
\hline
Within groups & 2 &  7.2365 & 3.61825 & 24.51 & $8.08 \times 10^{-11}$ \\
Between groups & 441 & 65.1077 & 0.14764 & & \\
Total & 443 & 72.3442 & & & \\
\hline
\end{tabular}
\caption{One-way ANOVA test results for the average pollen ring metric over day 20 to day 60. It is highly unlikely that the average pollen ring metrics from the simulation runs of models 1, 2, and 3 were drawn from distributions with the same mean. ANOVA tests for average brood clumping were even more pronounced.}\label{tab:anova}
\end{table}

\begin{figure}
\centering
\includegraphics[width=120mm]{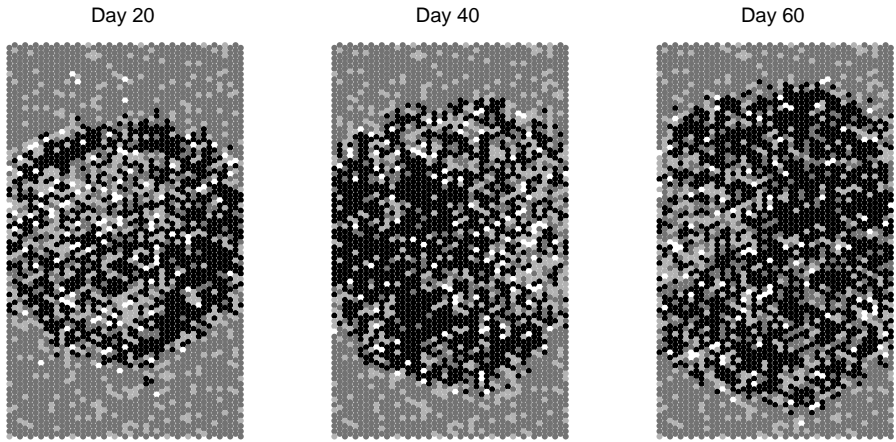}
\caption{Snapshots of the model 1 run which maximized the product of time-averaged metrics $\overline{m_b}\cdot \overline{m_p}$ at day 20, 40 and 60 in (a), (b), and (c), respectively. Cells containing brood are black; cells containing pollen are light gray; cells containing honey are dark gray; empty cells are white. $n=61,r_b=2,r_n=2,\omega=3475,\rho_{ph}=0.9638,\rho_p=0.9151,\rho_h=1.0668,\chi=1,k=10$.}
\end{figure}

\begin{figure}
\centering
\includegraphics[width=120mm]{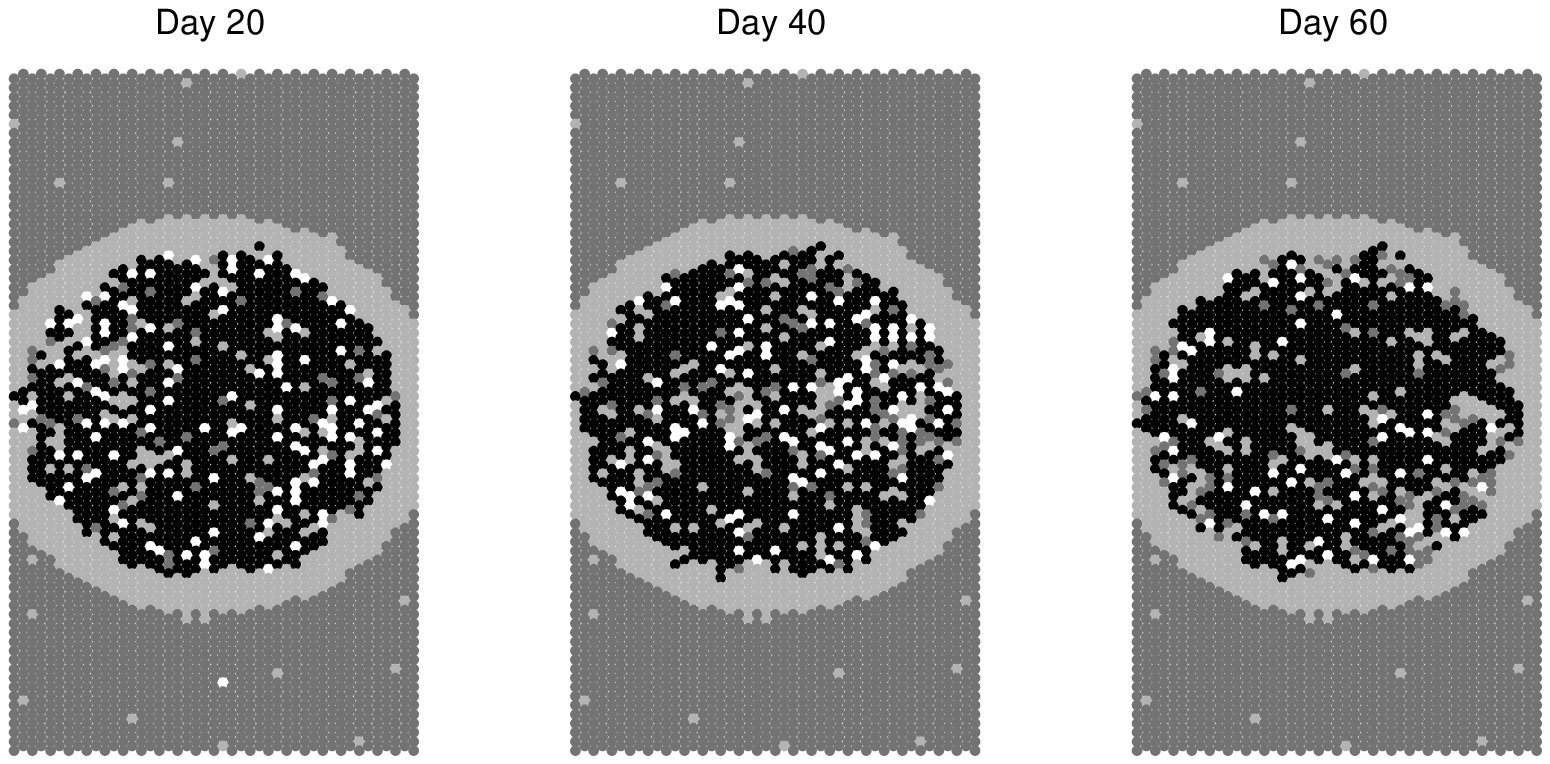}
\caption{Snapshots of the model 2 run which maximized the product of time-averaged metrics $\overline{m_b}\cdot \overline{m_p}$ at day 20, 40 and 60 in (a), (b), and (c), respectively. Cells containing brood are black; cells containing pollen are light gray; cells containing honey are dark gray; empty cells are white. $n=97,r_b=4,r_n=1,\omega=1150,\rho_{ph}=0.2764,\rho{r}_p=0.9362,\rho_h=1.0970,\chi=0,k=15$.}
\end{figure}

\begin{figure}
\centering
\includegraphics[width=120mm]{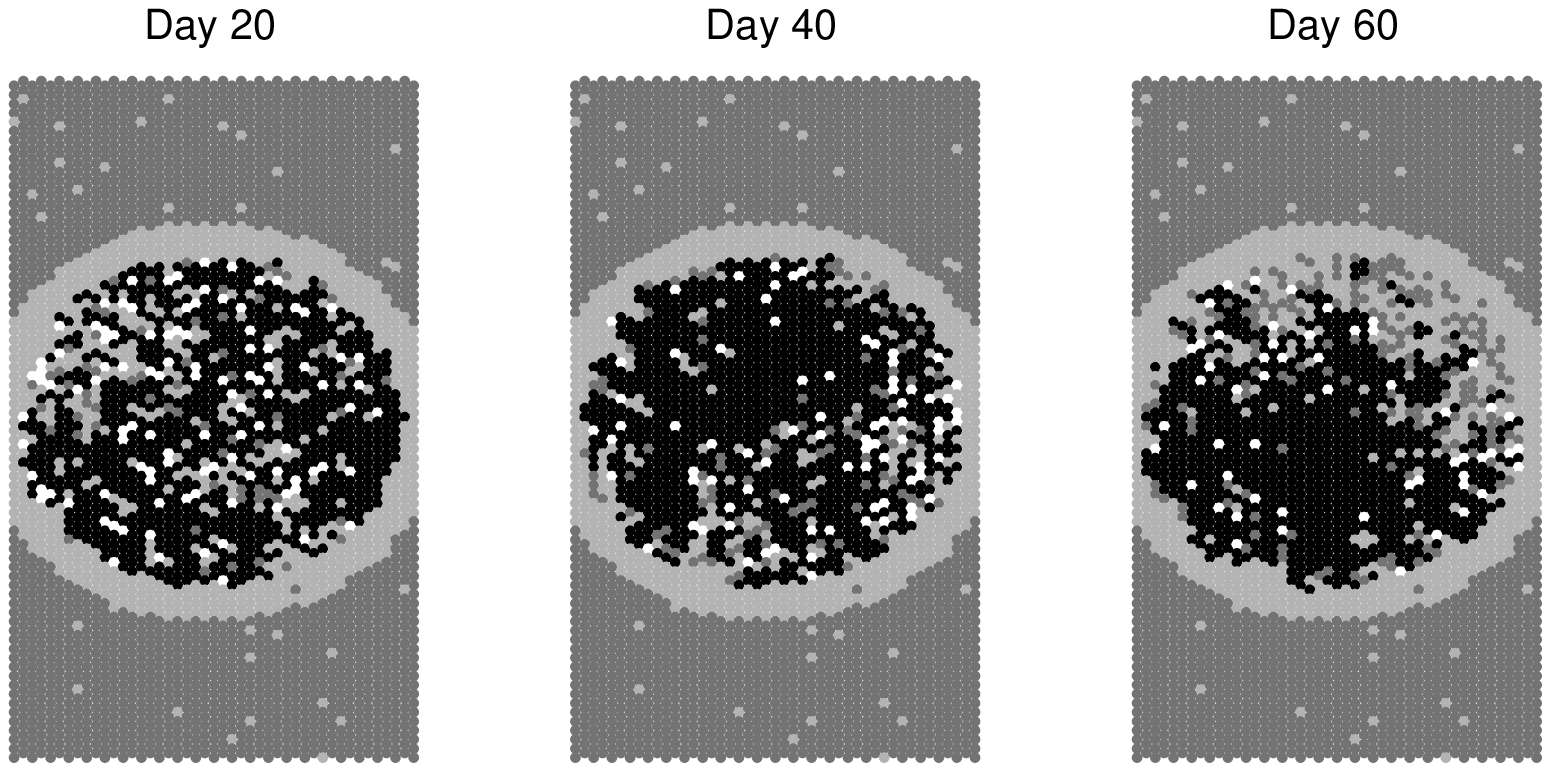}
\caption{Snapshots of the model 3 run which maximized the product of time-averaged metrics $\overline{m_b}\cdot \overline{m_p}$ at day 20, 40 and 60 in (a), (b), and (c), respectively. Cells containing brood are black; cells containing pollen are light gray; cells containing honey are dark gray; empty cells are white. $n=90,r_b=3,r_n=1,\omega=1000,\rho_{ph}=0.9719,\rho_p=1.0206,\rho_h=0.9854,\chi=0,k=16$.}
\end{figure}

\begin{figure}
\centering
\includegraphics[width=120mm]{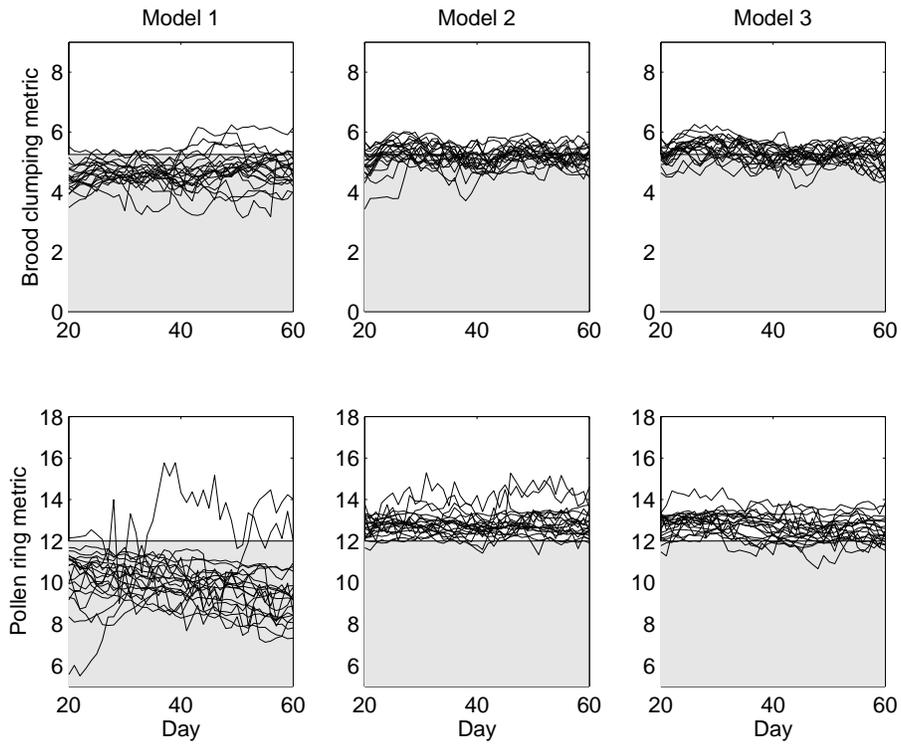}
\caption{Trajectories of the brood clumping metric and pollen ring metric the 30 best simulation runs for each model. Recall that if the brood clumping metric is above 5.25 and the pollen ring metric is above 12, then the pattern is considered to be well formed.}
\end{figure}

\begin{figure}
\centering
\includegraphics[width=120mm]{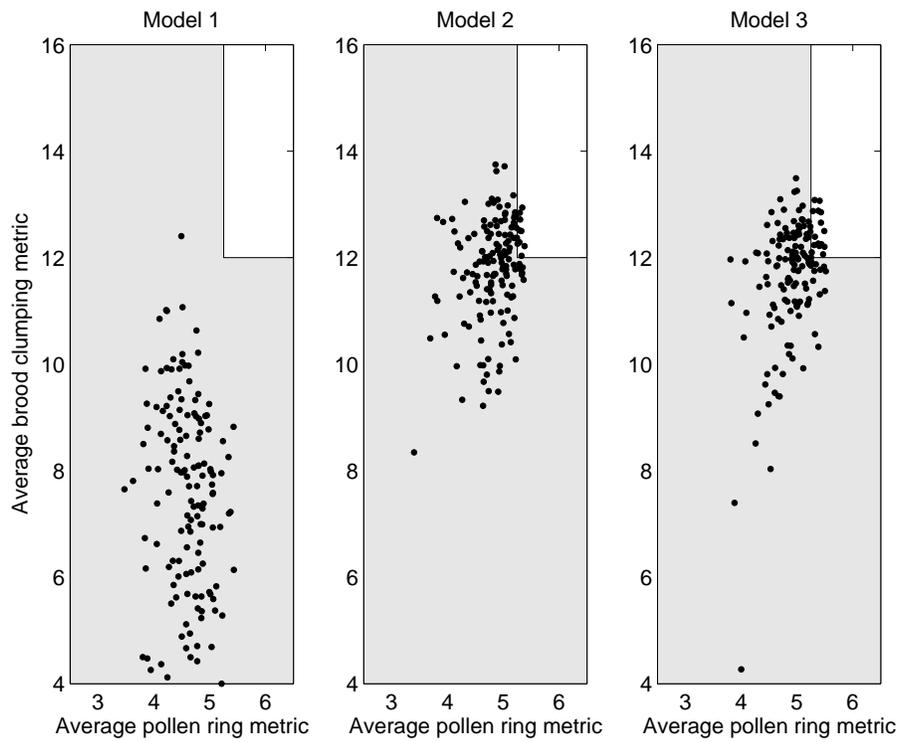}
\caption{Brood clumping and pollen ring metrics on day 60 for each of the 200 simulations of models 1, 2 and 3. The gray region represents metric combinations that result in a poorly defined cell allocation pattern.}
\end{figure}

\begin{figure}
\centering
\includegraphics[width=120mm]{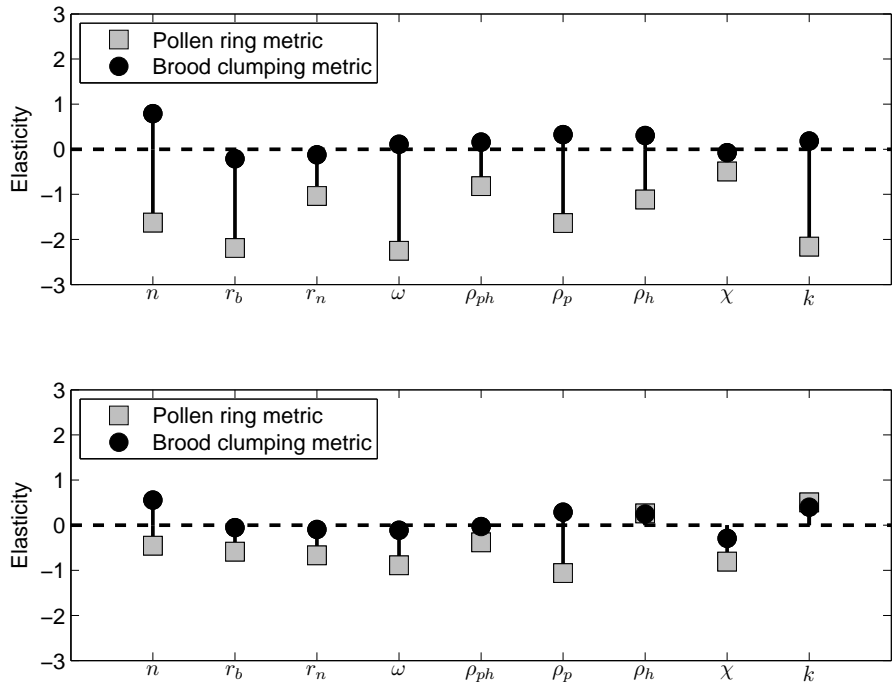}
\caption{Sensitivity metrics for models 1 and 2. The sensitivity of models 2 and 3 were qualitatively similar. The ``elasticity'' is the proportional change in the value of the metric relative to the change in the parameter value. A positive elasticity indicates that the metric increases with the parameter while a negative elasticity indicates that the metric decreases with increases in the parameter. See Table \ref{tab:param} for parameter definitions and ranges.}
\end{figure}

\begin{figure}
\centering
\includegraphics[width=120mm]{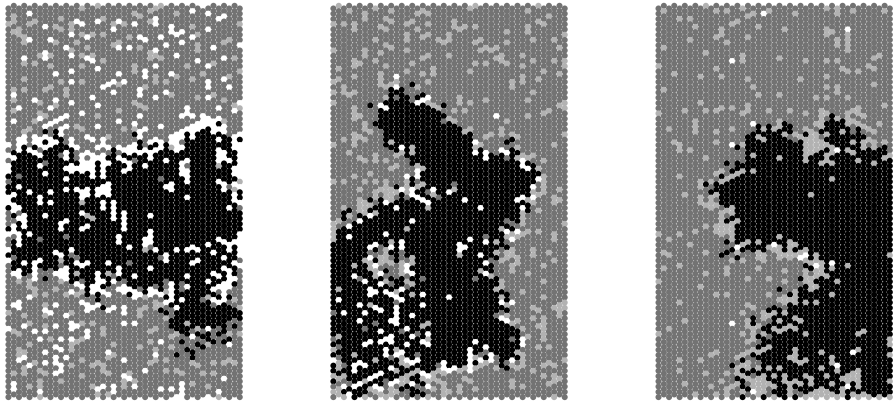}
\caption{A day 20 of a simulation run demonstrating that each model is able to create the desired pattern on an empty sheet of comb. Model 1 has parameters $n=79,r_b=3,\rho_h=3,\omega=1930,\rho_{ph}=0.2536,\rho_p=1.0436,\rho_h=0.6799,\chi=2,k=13$, model 2 used parameters $n=79,r_b=3,\rho_h=3,\omega=1930,\rho_{ph}=0.2536,\rho_p=1.0436,\rho_h=0.6799,\chi=2,k=13$, and model 3  used parameters $n=86,r_b=3,\rho_h=4,\omega=1480,\rho_{ph}=0.3748,\rho_p=1.0345,\rho_h=0.6900,\chi=2,k=10$.}
\end{figure}

\begin{figure}
\centering
\includegraphics[width=120mm]{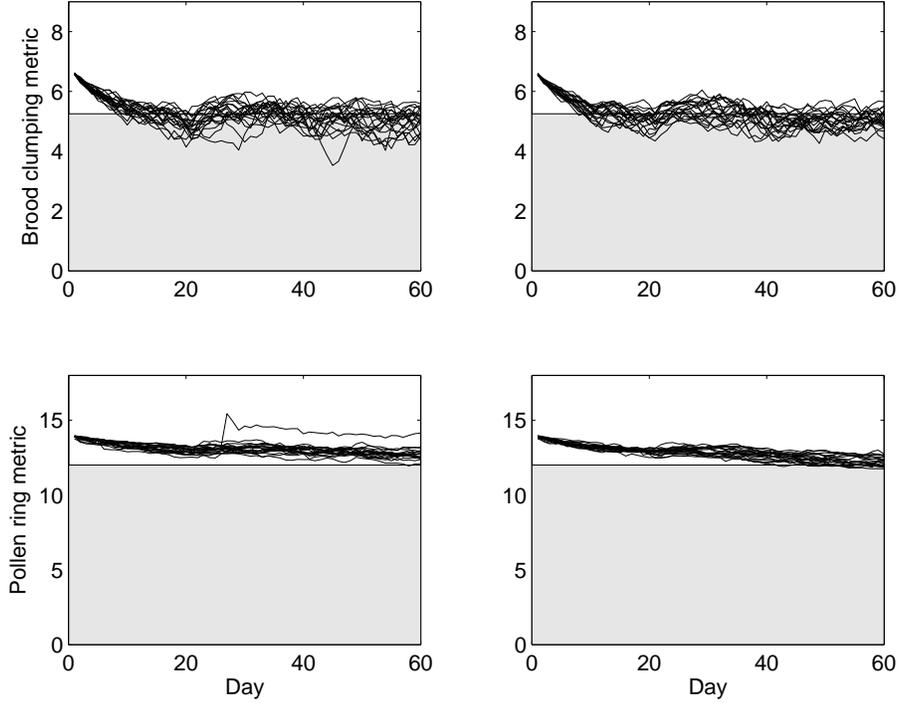}
\caption{To test the consistency of the model, 20 replicate simulations were performed for the parameter sets that resulted in the highest average metrics for models 2 (left panels) and model 3 (right panels). For model 2, this was $n=97,r_b=4,\rho_h=1,\omega=1150,\rho_{ph}=0.2764,\rho_p=0.9362,\rho_h=1.097,\chi=0,k=15$. For model 3: $n=90,r_b=3,\rho_h=1,\omega=1000,\rho_{ph}=0.9719,\rho_p=1.0206,\rho_h=0.9854,\chi=0,k=16$. Trajectories of the brood clumping metric (top panels) and pollen ring metric (bottom panels) for these twenty simulations show a tight fit for the pollen ring metric with a mean average metric (over the last 40 days) of $12.98\pm0.30$ S.D. for model 2, and $12.64\pm0.23$ S.D. for model 3 with all of the runs maintaining an average brood clumping metric above the desired $5.25$. The brood metric is more variable with a mean metric of $5.13\pm 0.17$ S.D.  for model 2, and $5.14\pm0.15$ for model 3. Only six model 2 and four model 3 runs had an average brood clumping metric above the desired $5.25$, but all of the runs had an average brood clumping metric above $4.79$.}
\end{figure}

\end{document}